\documentclass[fleqn,usenatbib]{mnras}

\usepackage{newtxtext,newtxmath}

\usepackage[T1]{fontenc}
\usepackage[dvipsnames]{xcolor}
\usepackage{soul}
\DeclareRobustCommand{\VAN}[3]{#2}
\let\VANthebibliography\thebibliography
\def\thebibliography{\DeclareRobustCommand{\VAN}[3]{##3}\VANthebibliography}

\usepackage{graphicx}	
\usepackage{stfloats}

\usepackage{amsmath}	

\usepackage{orcidlink}
\usepackage{ulem}

\title[Detection of 27 CBPs via apsidal precession]{Detection of 27 candidate circumbinary planets through apsidal precession of eclipsing binaries observed by \textit{TESS}}

\author[Thornton et al.]{
Margo Thornton\,\orcidlink{0009-0000-7387-2131}\thanks{E-mail: margo.thornton@unsw.edu.au},$^{1,2}$
Benjamin T. Montet\,\orcidlink{0000-0001-7516-8308},$^{1}$
Riley White,$^{1}$
Arden Shao\orcidlink{0009-0002-6875-5111},$^{3}$
 and Diya T. Kumar\orcidlink{0009-0004-1721-4803},$^{3}$
\\
$^{1}$School of Physics, University of New South Wales, Kensington NSW 2052, Australia\\
$^{2}$The SETI Institute,
339 N Bernardo Ave Suite 200, Mountain View, CA 94043, USA\\
$^{3}$Cahill Center for Astronomy and Astrophysics, California Institute of Technology,
1200 E. California Blvd., MC 249-17, Pasadena, CA 91125, USA
}

\date{Accepted 2026 March 11. Received 2026 March 10; in original form 2025 December 8}

\pubyear{\the\year{}}

\begin{document}
\label{firstpage}
\pagerange{\pageref{firstpage}--\pageref{lastpage}}
\maketitle

\begin{abstract}
Most circumbinary planets have been discovered by their transits, limiting our understanding of such systems to those with mutually coplanar architectures. This bias makes it difficult to infer the true circumbinary planet population, highlighting the need for alternative detection methods that do not rely on transits. In this work, we explore one such approach by leveraging apsidal precession as a dynamical signature of planetary companions. We analyse \textit{TESS} photometry of a sample of 1590 eclipsing binaries from the \textit{Gaia} DR3 Catalogue of Eclipsing Binary Candidates to identify systems exhibiting measurable apsidal precession that cannot be explained by general relativistic, tidal, or rotational effects alone. These excess precession signals point to the presence of additional gravitational perturbers and allow constraints to be placed on the masses and orbital separations of potential companions.  We present a new set of 27 candidate circumbinary planets identified through this precession-based method, as well as 6 candidate companions with a higher minimum mass. Their inferred properties remain degenerate, as the same dynamical signatures can arise from lower-mass planets at less than 1 AU or from more massive companions on wider, few-AU orbits, reflecting the current uncertainty in characterising these systems. Radial velocities can help break this degeneracy and provide direct confirmation.
\end{abstract}

\begin{keywords}
  methods: data analysis, observational -- eclipses -- exoplanets -- planets and satellites: detection -- binaries: eclipsing
\end{keywords}



\section{Introduction}

The existence of circumbinary planets, exoplanets that orbit binary stars, has been confirmed in just the last 15 years \citep{kepler16}. NASA’s \textit{Kepler} \citep{Kepler} and \textit{TESS} \citep{TESS} missions have enabled the discovery of 14 circumbinary planets (CBPs) via transits \citep[e.g.][]{kepler34, kepler38, kepler64}. All of these confirmed systems are transiting CBPs, detected because their orbital planes are closely aligned ($<5^{\circ}$) with that of the binary. These are highly reliable detections because of their distinctive transit signatures. However, relying primarily on transits may bias our current understanding of CBPs toward coplanar and compact systems, leaving a potentially large population of wider or misaligned planets undetected. Despite its all-sky coverage, \textit{TESS} has revealed only two CBPs \citep{toi1338b, TIC1729}, further illustrating the difficulty of detecting these planets through transits alone. The BEBOP Survey \citep{bebop} has uncovered 2 CBPs via the radial velocity (RV) method \citep{bebop1, bebop3}. Eclipse timing variations \citep[ETVs; ][]{kepler_1660}, gravitational microlensing \citep{microlensing}, and direct imaging \citep{squicciarini2025} have each led to widely accepted identifications of a circumbinary planet around a main-sequence binary. There is also a sample of more than a dozen proposed planets around post-common envelope binaries \citep[e.g.][]{lee2009, beuermann2010, qian2012} suggested through ETVs.

\begin{figure}
    \centering
    \includegraphics[width=1.0\linewidth]{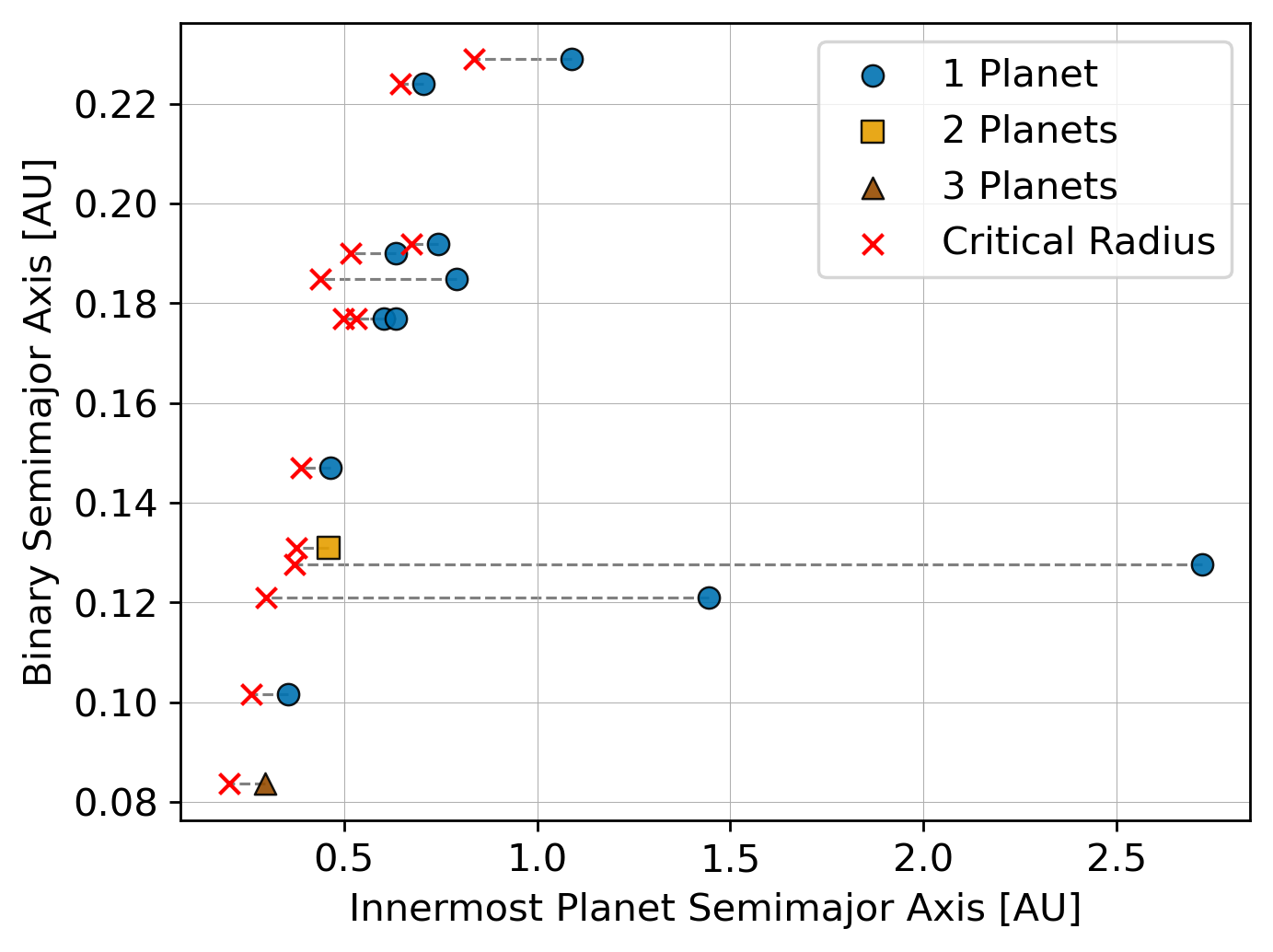}
    \caption{The distribution of the semimajor axis of the known CBPs around main sequence binaries with reference to the critical stability radius of the binary (denoted with ``x" connected by dashed line with data point). For the multiple planet systems, Kepler-47 \citep{kepler47} and TOI-1338 \citep{toi1338b, bebop1}, only the innermost planet is shown. The CBPs discovered via microlensing \citep{microlensing} and direct imaging \citep{squicciarini2025} have been omitted from this diagram due to their uncertain orbital parameters and large orbital separation, respectively. In a vast majority of the systems, the innermost planet exists just outside of the critical radius.}
    \label{fig:crit_rad}
\end{figure}

A vast majority of the known CBPs are coplanar with and exist just outside the critical stability radius \cite[$r_{\rm crit}$,][]{critical_radius} of their host binary, portrayed in Figure \ref{fig:crit_rad}; the critical radius defines the boundary for stable orbits. This radius, from \cite{critical_radius}, can be expressed as:

\begin{equation}
    r_{\rm crit} = a_{\rm in} (1.6 + 5.1e - 2.22e^2 + 4.12\mu - 4.27e - 5.09\mu^2 + 4.61e^2\mu^2)
\end{equation}

where $a_{\rm in}$, $\mu$, and $e$ are the binary's semimajor axis, mass ratio, and eccentricity, respectively. Theory predicts that CBPs preferentially reside near the inner edge of the circumbinary stability region \citep{cbp_rad, kley2019}. Separately, theoretical work also suggests that CBPs tend to evolve toward coplanar configurations with their host binaries \citep{cbp_incl}. Nevertheless, the search methods used so far are strongly biased toward detecting coplanar CBPs near the inner stability boundary. Transit surveys preferentially find planets whose orbits are nearly edge-on and aligned with the binary, because even a degree of mutual inclination dramatically reduces the probability of detecting repeated transits \citep{martin2015}. Similarly, planets orbiting just outside the stability radius produce frequent and more easily detectable transits, while planets at wider separations transit rarely or not at all over a mission baseline. Tests of the pile-up of CBPs near the critical radius are limited to transit and RV detections, as directly imaged CBPs are found at much wider separations and planets around post-common envelope binaries arise through fundamentally different formation and evolutionary pathways.  These selection effects, along with the small sample size, emphasise the importance of developing new methods to find CBPs. 

These observational biases are especially problematic because theoretical predictions for circumbinary planet formation are widely divergent. Some models suggest that circumbinary disks damp inclinations and produce highly coplanar planets \citep{cbp_incl}. Others predict that the binary can disturb the disk, causing it to warp or become turbulent, which may tilt planetary orbits or suppress planet growth \citep{abod2022, franchini2019, chen2019}. Without a representative sample, including misaligned or wide-orbit CBPs, we cannot distinguish between these competing scenarios, nor can we assess whether planet formation around binaries is more or less efficient than around single stars. This question has broader implications: microlensing surveys, which have found many free-floating planets, predict that they may originate from circumbinary disks through dynamical ejection \citep[e.g.][]{sutherland2016, Fitzmaurice2022, coleman2024, chen2024}, meaning that the efficiency of circumbinary planet formation contributes to the overall population of unbound planets. Altogether, these uncertainties highlight the need for less biased, dynamically informed detection methods capable of revealing the full diversity of circumbinary planetary systems.

In \cite{TIC1729}, the authors present evidence that the argument of periastron, $\omega$, of the binary star is precessing over time, a motion attributed to the gravitational influence of a CBP that was initially detected via its transits. Here, $\omega$ denotes the angle between periastron and the reference plane (ascending node), which is measured in the direction of the orbiting star's motion within the orbital plane. This effect is known as apsidal precession (due to the motion of the line of apsides), and can be caused by a third body in the system perturbing the orbit of the inner binary \citep{harrington1968}. Because apsidal precession corresponds to a nonzero $\dot{\omega}$, it leads to a time-varying geometry of the orbit. In an eccentric binary, the relative timing of the primary and secondary eclipses encodes the quantity $e\cos\omega$ \citep[e.g.][]{hilditch2001}, so a changing $\omega$ causes $e\cos\omega$ to vary as well. This variation produces a measurable shift in the interval between successive primary and secondary eclipses, providing a direct way to detect apsidal precession. 

It is important to distinguish apsidal precession from nodal precession, which refers to the precession of the orbital plane itself about the system’s invariable plane. Nodal precession changes the orbital inclination relative to the line of sight and is responsible for bringing CBPs in and out of transit, making it the dominant effect discussed in much of the CBP literature to date \citep[e.g.][]{schneider1994, kostov2014}. In contrast, apsidal precession affects the orientation of the orbit within its plane and presents as timing variations rather than changes in transit visibility. While these two effects are physically distinct, their characteristic timescales are often comparable in circumbinary systems \citep{leung2013}.

Here, we focus on the apsidal precession of the inner binary, which directly affects the relative timing of the primary and secondary eclipses. The phase difference between the primary and secondary eclipses can be represented using $\omega$. If $\dot{\omega} \neq 0$, the primary and secondary eclipse times drift away from a strictly periodic ephemeris, resulting in a pattern where the primary eclipses and secondary eclipses diverge. With enough data covering a long time interval, this divergence becomes roughly sinusoidal, where a full apsidal period is the time it takes for $\omega$ to change by $360^\circ$ \citep{dimoff2023}. For the kinds of systems considered here, the apsidal periods are typically on the order of decades. For example, the system analysed by \cite{TIC1729} has an apsidal period of roughly 50 years.


A distant companion is not the only possible cause of precession to a binary star. Effects due to general relativity (GR) and the tides and rotation of the stars can contribute to the precession \citep[e.g.][]{einstein1916, cowling1938, sterne1939}. These effects are more prominent for high-mass and short-period binaries, respectively. Importantly, the magnitudes of GR and tidal/rotational precession can be calculated precisely when the stellar and orbital parameters are known \citep[e.g.][]{gimenez1985}. This allows the expected precession from these mechanisms to be separated from any additional precession induced by a third body in a wider orbit, which we explore in more detail in the analysis that follows.

Apsidal precession is a common dynamical phenomenon observed in a wide range of systems. Many hierarchical stellar triples show measurable apsidal motion driven by the outer companion \citep{mazeh1979}, and even within our own Solar System, Mercury’s orbit precesses at a rate of 43 arcseconds per century due to general relativity \citep{einstein1916} and precesses because of other bodies in the Solar System by 500 arcseconds per century. These examples highlight how sensitive apsidal motion is to external perturbations, making it a powerful tool for identifying and characterising CBPs.

In this work, we use the detection of apsidal precession in a binary to infer the presence of a third body orbiting the inner binary. Historically, this technique has been difficult to apply because long baselines of precise, continuous eclipse timing have rarely been available. \textit{TESS} now provides multi-year, high-cadence photometry for tens of thousands of eclipsing binaries, enabling the detection of coherent apsidal motion that would have been unmeasurable in earlier surveys. Given that \textit{TESS} has found only two CBPs via transits, exploiting its time-domain precision to search for dynamical signatures such as apsidal precession offers a timely and complementary pathway for expanding the CBP population.

In the sections that follow, we outline our approach to identifying and characterising precessing eclipsing binaries. Section \ref{sec:methods} describes the data we use (\S\ref{subsec:data}), the data processing steps (\S\ref{subsec:processing}), our determination of precise binary periods (\S\ref{subsec:periods}), the extraction of eclipse times (\S\ref{subsec:eclipses}), and refining the EB sample for precession analysis (\S\ref{subsec:special}). We then present our calculation of apsidal precession rates (\S\ref{subsec:precession}) and the resulting constraints on potential third bodies (\S\ref{subsec:body}). Section \ref{sec:results} introduces the sample of circumbinary planet candidates uncovered by this method, and in Section \ref{sec:discussion} we discuss their implications and prospects for future work before concluding in Section \ref{sec:conclusion}.

\section{Methods} \label{sec:methods}

\subsection{Data} \label{subsec:data}

Starting with the \textit{Gaia} DR3 catalogue of eclipsing binary candidates \citep{Gaia}, we select a sample of targets to investigate for signs of precession. The initial selection criteria are as follows: parallax $> 1 \: mas$, $m_{\rm G} < 18$, non-ellipsoidal, main-sequence stars, and available \textit{TESS} light curve data (with at least a two year baseline). From here, we apply the following analysis to the EBs with processed light curve data from MIT's Quick-Look Pipeline \cite[QLP;][]{qlp}. 

\subsection{Data Processing} \label{subsec:processing}

For each target, we retrieve all available \textit{TESS} QLP light curves using \texttt{lightkurve} \citep{lightkurve2018}. We use only data points without any quality flags applied (i.e. \texttt{QUALITY}\,=\,0), ensuring that our analysis is restricted to measurements free of known spacecraft or pipeline anomalies. This conservative filtering avoids contamination from scattered light, cosmic rays, or other events that could mimic or distort eclipse features. After stitching the downloaded light curves across all available sectors, we assess the background level of each datum using the \texttt{sap\_bkg} values. We remove data points whose background is more than $3\sigma$ from the mean, which reduces contamination effects. Next, we remove flares and other positive outliers using one-sided sigma clipping. We do not apply lower clipping in order to preserve the eclipses.

To detrend, we obtain an initial estimate of the binary period and eclipse properties, which are required to construct reliable eclipse masks. We first apply a Box-Least Squares \cite[BLS;][]{kovacs2002} search to the full light curve to obtain a coarse estimate of the dominant periodicity. Because BLS periodograms are prone to aliasing, we apply an automated correction procedure to the BLS period solution. To implement this correction, we use the \texttt{compute\_stats} function of the periodogram. From this we extract \texttt{depth}, \texttt{depth\_half}, \texttt{depth\_odd}, and \texttt{depth\_even}, which represent the inferred eclipse depth at the reported period, half the period, twice the period, and twice the period with a phase offset of one orbital period, respectively. If the depth at half the reported period is within 5\% of the primary depth, the signal is likely a harmonic at 2 times the true period and we adopt half the reported period. Conversely, if \texttt{depth\_odd} and \texttt{depth\_even} are nearly equal (within 5\%) but differ significantly from the primary depth, this suggests a misidentified period at half the true value, and we adopt twice the reported period. Otherwise, we retain the original period.

Considering the primary and secondary eclipses can have similar depths, we perform an additional verification step by counting the number of distinct eclipse minima on the phase-folded light curve. If more than two minima are detected, the period is halved; if fewer than two are detected, the period is doubled. This ensures that the adopted period produces exactly one primary and one secondary eclipse per orbital cycle.

Using this alias-corrected period as an initial guess, we then run \texttt{transitleastsquares} \cite[TLS;][]{tls} on a tighter period constraint (0.99 to 1.01 times the alias-corrected BLS period). TLS returns a model with refined estimates of the eclipse ephemeris and duration. We construct eclipse masks using the TLS-derived duration, expanding each eclipse window to 3.5 times the reported duration to ensure complete coverage of the eclipse and surrounding ingress/egress. We visually inspected this mask prescription for approximately 50 systems and found that it consistently enclosed the full eclipse with ample margin.

Because QLP systematics are strongly sector-dependent, we flatten each sector individually. Using the initial period and eclipse masking method described above, we (1) mask the primary eclipses and perform a BLS search on the remaining data to identify the secondary eclipses, (2) construct a combined eclipse mask, and (3) apply a sliding-window flattening procedure that removes long-term trends while preserving the eclipses.

Flattening each sector independently avoids injecting discontinuities at sector boundaries and ensures consistent baseline normalisation before phase folding. Finally, we concatenate the flattened sectors to produce a single detrended light curve.

\subsection{Finding Periods} \label{subsec:periods}

In order to detect any variations in the eclipse timing, we must find precise periods of the primary and secondary eclipses independently. We therefore apply a similar eclipse-masking strategy to that used for detrending, but we now use the refined period. The primary eclipse is identified in the phase-folded light curve from the deepest flux minimum. Its width is estimated from the half-depth points during ingress and egress, and a conservative mask is constructed that extends several eclipse widths beyond the measured duration. This mask is used to remove the primary eclipse from the light curve, and TLS is then applied to the remaining data within a narrow window around the initial period estimate to refine the period of the secondary eclipse. The procedure is then repeated in reverse, masking the secondary eclipses to refine the period of the primary. By iteratively masking and re-fitting each eclipse sequence, we ensure that the two signals are analysed independently. This prevents contamination between the primary and secondary eclipses and enables precise measurements of their respective ephemerides, which is essential for identifying apsidal precession.

Any deviation in these two periods would indicate precession. For EBs with periods shorter than 3 days, we mark it as ``short-period" and refrain from calculating precession effects of that binary. Tidal effects scale as $(a/R)^5$, where $a$ is the semimajor axis of the binary and $R$ is the radius of a given star in the system; thus, short-period EBs are more prone to tidal and rotational effects, and their light curves often entail more variation.

\begin{figure*}
    \centering
    \includegraphics[width=1.0\linewidth]{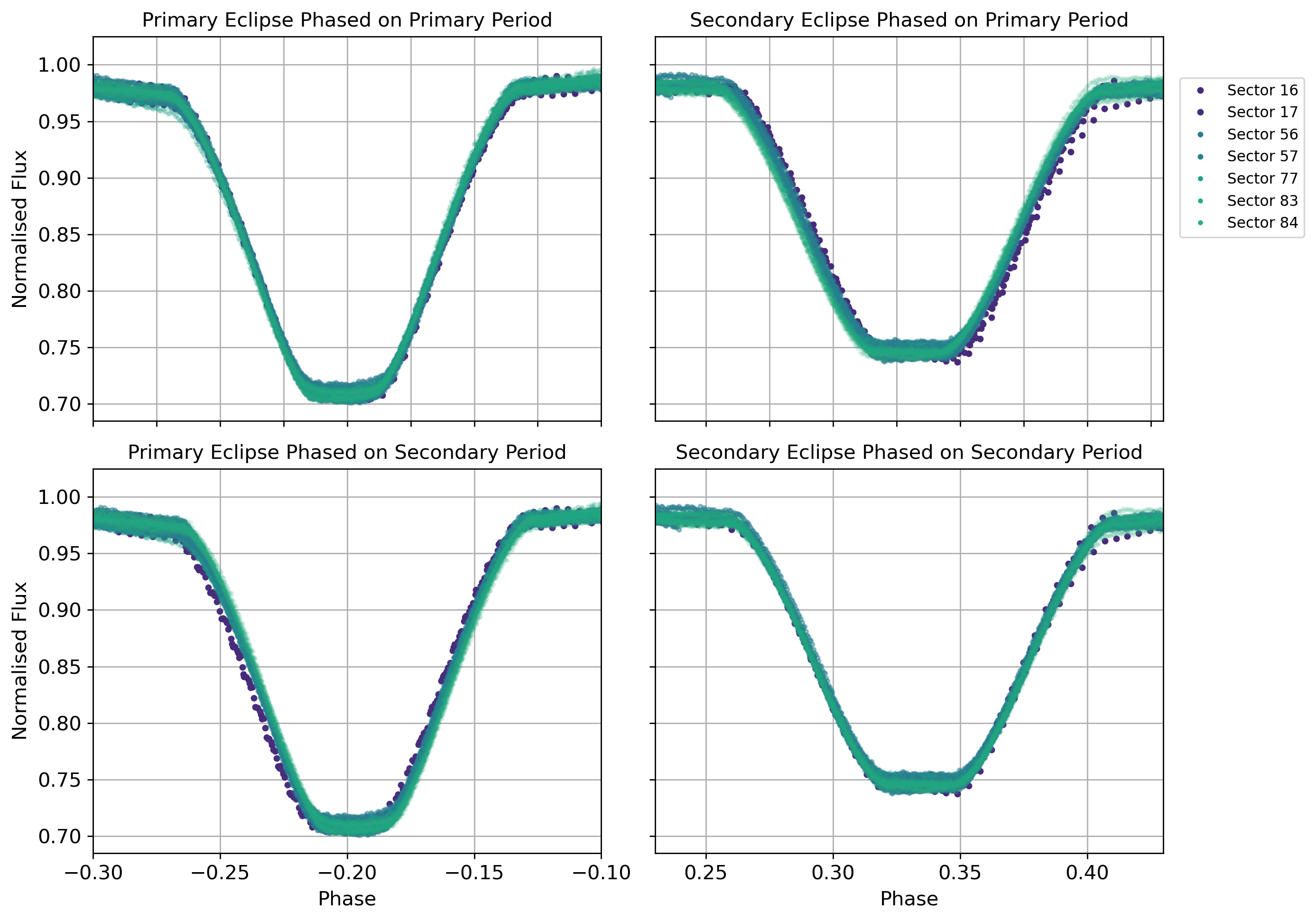}
    \caption{The phase-folded light curve of TIC 343127696, a $\sim4.67$-day eclipsing binary, showing seven sectors ($\sim 5$ years) of data. The primary eclipses are in the left two panels, while the secondary eclipses are in the right two panels. On top, the light curve has been folded on the period of the primary eclipses while on the bottom, it has been folded on the period of the secondary eclipses. When folded on the primary period, only the primary eclipses align cleanly; when folded on the secondary period, only the secondary eclipses align clearly. This systematic offset between the two eclipse types demonstrates that their inferred periods differ, which is direct evidence of apsidal precession in the binary.}
    \label{fig:TIC343127696_4panel}
\end{figure*}

To make sense of the two period values, we can visually inspect the folded eclipses on both periods to detect any variation, as seen through an EB in our sample in Figure \ref{fig:TIC343127696_4panel}. Precession will appear if the primary eclipse phased on the primary period lines up well, whereas the secondary eclipse phased on the same period does not: the secondary eclipses will appear to occur earlier or later in time. The opposite effect will occur for the light curve folded on the secondary period. In addition to detecting precession, this step provides several auxiliary benefits. It allows us to verify that the derived periods are correct and also reveals astrophysical features such as flares, variable eclipse depths, starspot modulation, or out-of-eclipse variability. While visual inspection may not always be sufficient to catch signs of precession, especially when looking through a large sample, it serves as an essential early diagnostic to validate the robustness of the period determination before proceeding to more quantitative analyses.

\subsection{Measuring Eclipse Times} \label{subsec:eclipses}

In addition to looking for visual cues of precession in the light curve, we must also quantitatively analyse the mid-eclipse times to characterise the structure of the ETVs, determining whether they follow a simple linear trend or exhibit more complex variations indicative of apsidal precession or other dynamical effects. 

We calculate the eclipse times using four different methods. The first method obtains the time at half-depth of each eclipse during ingress and egress, here defined by everything within $0.1 P_{\rm Binary}$ to the left and right of the mid-eclipse time determined by TLS, and uses the midpoint of the two as the mid-eclipse time. This approach is simple and works well for symmetric eclipses with minimal noise, but can be biased if the eclipse shape is distorted. The second method folds each eclipse at trial midpoints. We determine the mid-eclipse time as the midpoint that best aligns ingress and egress when folded. This technique is particularly useful when small variations in eclipse depth or shape exist between cycles. The third method models the phase-folded primary and secondary eclipse with SciPy's \texttt{LSQUnivariateSpline} interpolator \citep{scipy}. We compare the eclipse model to each corresponding eclipse in the dataset. By sliding the model along the eclipse at small increments, we determine the shift that best matches the data to the model. This method is effective at capturing subtle shifts in eclipse timing even in the presence of noise or asymmetries caused by starspots or other stellar activity. The fourth method is a similar process to the previous, but instead of modelling the full eclipse, we only model the ingress and egress, then take the midpoint of the best-fitting shift for each model. This approach is advantageous when the eclipse bottoms are distorted or affected by depth variations, as it relies on the more stable ingress and egress regions rather than the variable central portion of the eclipse. By employing multiple methods, we can cross-check eclipse times, identify outliers, and ensure robust measurements under a variety of observational and astrophysical conditions. To ensure reliable timing measurements, we exclude eclipses that are poorly sampled. We define ingress and egress here as the portions of the eclipse spanning $10-90\%$ of the total eclipse depth on either side of the eclipse minimum, and we require a minimum of three data points in each region for an eclipse to be included. This criterion reduces susceptibility to timing biases from sparse coverage, gaps, or outliers, and ensures that the eclipse morphology is sufficiently resolved.

For this initial search, we do not directly fit physical eclipse models (e.g., PHOEBE \citep{prsa2018} or starry \citep{luger2019}), as these approaches are computationally expensive and impractical for our large initial eclipsing binary sample. Instead, we adopt model-independent timing estimators that enable rapid and robust eclipse-time measurements across thousands of systems. We plan to apply more detailed physical modelling to promising candidates in future work to better characterise their system parameters and dynamical properties.

To calculate uncertainty on the fitted eclipse midpoint, we use \texttt{lmfit} \citep[Non-Linear Least-Squares Minimization and Curve-Fitting for Python;][]{lmfit, Levenberg1944, Marquardt1963}. This tool uses a constrained least-squares optimisation. We first define the eclipse time as a free parameter within an allowed interval centered on the predicted mid-eclipse time, ensuring that the fit remains physically plausible while still permitting deviations driven by the data. The parameter and its bounds are passed to the \texttt{lmfit} minimiser, together with the residual function and the observational time series, where the flux uncertainty is approximated by the standard deviation of the out-of-eclipse data. The minimiser then performs a Levenberg–Marquardt \citep{Levenberg1944, Marquardt1963} least-squares fit, varying only the eclipse time to identify the value that minimises the residuals between the observed and modelled light curve. Upon convergence, we calculate the covariance matrix of the solution with \texttt{lmfit}, from which the formal 1$\sigma$ uncertainty on the eclipse time is derived. This value, reported as \texttt{stderr}, represents the propagated uncertainty derived from the local curvature of the $\chi^2$ function at the best-fitting parameter value. We consider the uncertainties determined with this method reasonable because they account for the observed scatter in the out-of-eclipse flux and reflect the sensitivity of the fit to shifts in eclipse timing. Tests with synthetic and real data show that \texttt{stderr} closely matches the scatter in recovered mid-eclipse times, giving us confidence that the reported uncertainties accurately represent the true timing precision.

\begin{figure}
    \centering
    \includegraphics[width=1.0\linewidth]{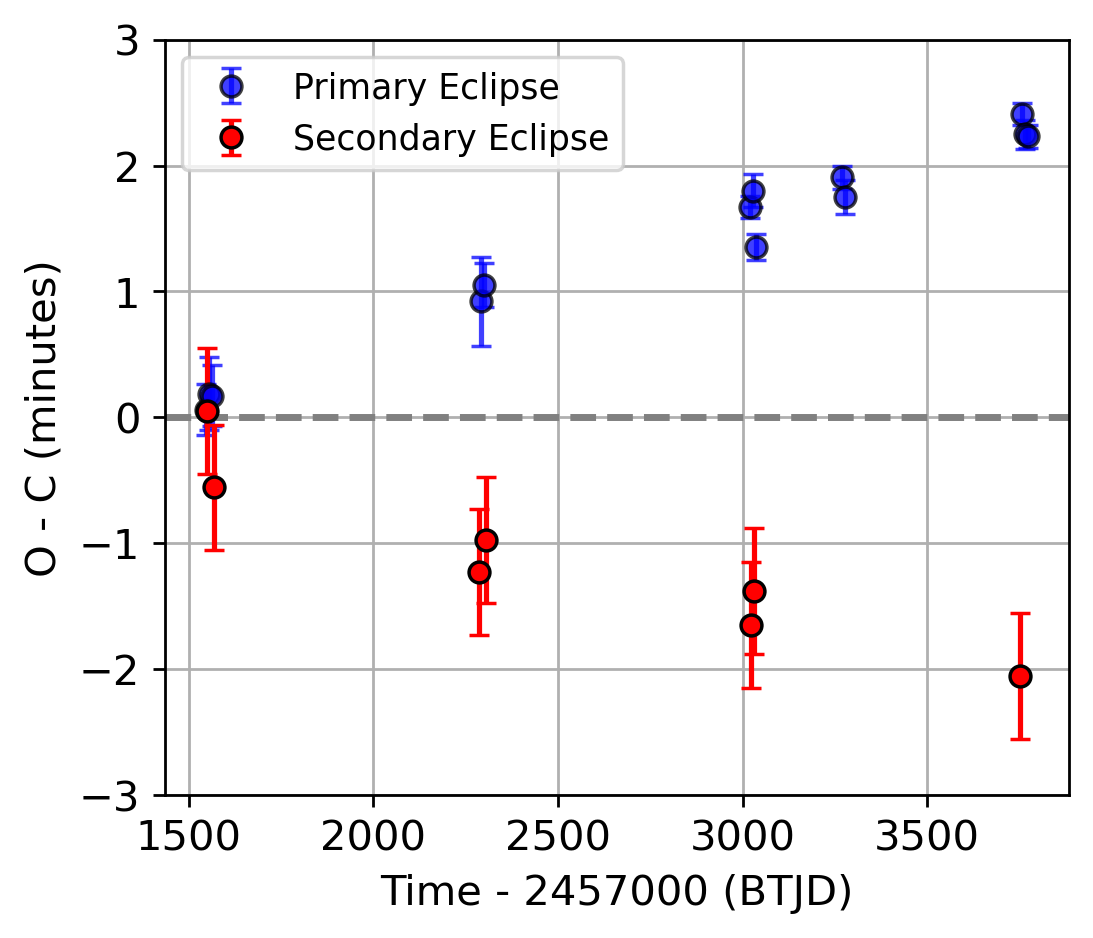}
    \caption{The observed minus calculated (O$-$C) plot based on a common period for TIC 286310830, a $\sim8.37$-day eclipsing binary. The primary eclipses are occurring later over time while the secondary eclipses occur earlier than expected. This divergence in eclipse timing is a sign that the system is precessing.}
    \label{fig:TIC286310830_OCwErr}
\end{figure}

With the observed eclipse times, we compare each with the expected eclipse time calculated on the basis of a common period (the average of the primary and secondary eclipse periods). The difference between these two values, the Observed minus Calculated (O$-$C), is shown for the primary and secondary eclipses of an EB in our sample in Figure \ref{fig:TIC286310830_OCwErr}. Because we determine eclipse times using four different methods, we select the method for each system that minimises the scatter in the O-C eclipse times. Specifically, we use the method that produces the lowest weighted root-mean-square (RMS) of the O$-$C residuals. If the inferred period of the primary and secondary eclipses based on these eclipse times are different and we see two diverging lines, as in Figure \ref{fig:TIC286310830_OCwErr}, we flag it as a precessing system. If one eclipse type appears to occur earlier over time while the other appears to occur later at relatively the same rate, we know the argument of periastron must be precessing.

 \subsection{Refining the EB Sample for Precession Analysis} \label{subsec:special}

The initial cuts described in Section \ref{subsec:data} reduce the full set of $\sim2$ million \textit{TESS} EBs to 12,819 systems suitable for further analysis. To identify those amenable to precession measurements, we apply several additional filters. First, because the typical EB period is $\sim2.5$ days \citep{kepler_ebs}, many systems fall below the timescale where dynamical precession from a third body can be cleanly distinguished from tidal and rotational effects. We therefore exclude all binaries with orbital periods shorter than 3 days, removing 7,904 systems and leaving 4,915 longer-period EBs. Next, we require a minimum of two primary and two secondary eclipses to obtain a reliable precession rate and to avoid aliasing in the O$-$C signal. Also, automated period and eclipse timing determinations fail in cases due to instrumental and astrophysical variabilities in the light curves. To identify these systems, we perform a visual inspection to confirm the cause. EBs are excluded when their light curves display large out-of-eclipse modulations (e.g. strong starspots, pulsations, flares, ellipsoidal variation,  instrumental noise, contamination) that prevent a stable and repeatable mid-eclipse measurement. This step relies on by-eye vetting because the variability morphologies are diverse and not well captured by a single quantitative metic. A total of 3,325 EBs fail these requirements.

After applying these criteria, 1,590 EBs remain as our ``clean" sample for initial O$-$C inspection. From these systems, we identify 71 binaries that show clear signatures of apsidal precession, for which we perform the full precession rate calculation and dynamical inference presented in \ref{subsec:precession} and \ref{subsec:body}.

 \subsection{Calculating Precession} \label{subsec:precession}

From the pool of precessing EBs, we quantify the precession, following \cite{hilditch2001} and \cite{charbonneau2005}, starting with the calculation of $e \cos{\omega}$:

\begin{equation} \label{eq:ecosomega}
    e \cos{\omega} = \frac{\pi}{2P} (t_{\rm I} - t_{\rm II} - \frac{P}{2} + N\Delta P)
\end{equation}

where $P$ is the average period of the eclipses, $t_{\rm I}$ is a mid-primary eclipse time, $t_{\rm II}$ is a mid-secondary eclipse time, and $N$ is the number of cycles between the eclipses. We calculate this value with two consecutive eclipses, so the last term goes to zero. Next, we calculate $e \sin{\omega}$, again following \cite{hilditch2001} and \cite{charbonneau2005}:

\begin{equation} \label{eq:esinomega}
    e \sin{\omega} \approx \frac{\Theta_1 - \Theta_2}{\Theta_1 + \Theta_2}
\end{equation}

where $\Theta_1$ and $\Theta_2$ are the widths of the primary and secondary eclipses, respectively. The widths are measured by eye based on the total duration of the eclipse from the start of ingress to the end of egress. This expression implicitly assumes that eclipse durations depend primarily on orbital geometry through $\omega$, but in practice they also depend on the impact parameter (or inclination) of each eclipse. Because we do not model the impact parameter explicitly, our estimates of $e\sin{\omega}$ carry additional systematic uncertainty. As a result, $e\cos{\omega}$ is significantly better constrained than $e\sin{\omega}$, which we highlight as the dominant source of uncertainty in the derived eccentricities and arguments of periastron. With the results of equations \ref{eq:ecosomega} and \ref{eq:esinomega}, we calculate the eccentricity, $e$. Using equation \ref{eq:ecosomega} at two different orbital cycles (typically the first and last observed), we measure $\frac{\text{d}}{dt} \left (e\cos{\omega} \right )$, which can also be written as

\begin{equation}
    \frac{\text{d}}{dt} \left (e \cos{\omega} \right ) = -e \dot\omega\sin{(\omega)}. 
\end{equation}

Rearranging gives

\begin{equation}
    \dot\omega = \frac{\text{d}}{dt} \left ( \cos{\omega} \right ) \frac{1}{\sin{\omega}}.
\end{equation}

Here, we assume that the eccentricity $e$ remains approximately constant over the \textit{TESS} observational baseline. This is generally a good approximation for hierarchical triples, where the apsidal precession timescale is much shorter than the eccentricity modulation timescale. However, perturbations from a third body can in principle induce measurable changes in $e$ over multi-year baselines \citep[e.g.][]{mckee2023}, so this assumption introduces an additional (usually small) systematic uncertainty.

The measured $\dot\omega$ therefore reflects the combined contributions from GR, tides and rotation, and any additional precession induced by a third body, all of which are additive. Here, we explicitly assume that all of these contributions act in the same (prograde) direction. This assumption is valid for the GR and tidal/rotational terms, and also for third-body perturbations, provided the system is not in a Kozai-Lidov \citep{kozai1962, lidov1962} regime. In that case, the argument of periastron can exhibit more complex behaviour, like libration, rather than steady apsidal circulation. Thus, the precession due to a third body would be the difference between the total observed precession and that induced by tides and GR:

\begin{equation}
    \dot\omega_{\rm 3rd} = \dot\omega - \dot\omega_{\rm GR} - \dot\omega_{\rm CL}
\end{equation} 

where $\dot\omega_{\rm GR}$ and $\dot\omega_{\rm CL}$ are the precession due to GR and classical (tidal and rotational) effects, respectively. The precession due to GR, following \cite{gimenez1985} and \cite{dimoff2023}, can be written as:

\begin{equation}
    \dot\omega_{\rm GR} = (5.447127276 \times 10^{-4}) \left( \frac{1}{1 - e_2} \right) \left ( \frac{M_1 + M_2}{P} \right ) ^{\frac{2}{3}} \text{ deg cycle}^{-1} .
\end{equation}

$M_1$ and $M_2$ are the masses of the primary star and secondary star, respectively, in units of solar masses, and $P$ in units of days. As we only select systems which have been identified as main-sequence binaries, we use temperature values from \cite{stassun2019} and temperature-mass relations from \cite{tmr} to determine the masses of the stars in the binary. Precession due to tidal and rotational effects, following \cite{baroch2021}, can be represented by:

\begin{equation}
    \dot\omega_{\rm CL} = 360 \times \sum_{i=1}^{2} \left (k_i c_i^{\rm rot} + k_i c_i^{\rm tid} \right ) \text{ deg cycle}^{-1}
\end{equation}

where $k_i$ is the internal structure (or apsidal) constant \citep{claret1993}, which describes how mass is distributed inside a star; a star with centrally concentrated mass has a small $k$ (<0.02), whereas a star with a more uniform mass distribution has a larger $k$ (>0.1). We adopt $k_i = 0.01$ as is used in the work of \cite{TIC1729}, adapted from the work of \cite{kterm} as an average value of the term for main sequence stars. This choice lies toward the high end of the range reported by \cite{claret1993}, which spans $\log k_i \sim -3$ to $-1.9$, corresponding to $k \lesssim 0.013$. Adopting this upper-end value instead of our $k_i = 0.01$ would increase $\dot\omega_{\rm CL}$ by at most a factor of $1.3$.  In most cases, the tidal and rotational contributions are the smallest component of the total precession budget, so this uncertainty does not significantly affect our calculations.

The rotational term, $c_i^{\rm rot}$, following \cite{baroch2021}, is calculated as:

\begin{equation}
    c_i^{\rm rot} = \frac{r_i^5}{(1 - e)^2} \left ( 1 + \frac{M_{3-i}}{M_i} \right ) \left ( \frac{\Omega_i}{\Omega_{\rm m}} \right )^2
\end{equation}

$\Omega_i$ and $\Omega_{\rm m}$ are the angular velocity of the star’s rotation and the angular velocity of its orbital
motion, respectively. Tidal theory and observations both show that binaries with periods $\lesssim10-12$ days are typically synchronised \citep{zahn1977, meibom2005}. We therefore assume $\Omega_i = \Omega_{\rm m}$. $r_i$ is the ratio of the star’s radius to the semimajor axis of its orbit, where the radius is determined using temperature-radius relations from \cite{tmr} and the semimajor axis is determined through Kepler's 3rd law and our precise determination of the binary's orbital period. The tidal term, $c_i^{\rm tid}$, following \cite{baroch2021}, can be expressed as:

\begin{equation}
    c_i^{\rm tid} =15 r_i^5 \frac{M_{3-i} \left (1 + 1.5e^2 + 0.125e^4 \right )}{\left (1 - e^2 \right )^5}
\end{equation}

A positive $\dot\omega_{\rm 3rd}$ value indicates that there is precession consistent with an additional perturbation caused by a third body in the system orbiting the EB. We classify the remaining systems with a positive $\dot\omega_{\rm 3rd}$ as candidate EBs to host a third body companion.

\subsection{Defining Parameter Space of Third Body} \label{subsec:body}

The value of $\dot\omega_{\rm 3rd}$ allows us to determine the permitted parameter space for the mass and semimajor axis of the third body, following \cite{miralda2002} using:

\begin{equation}
    \dot\omega_{\rm 3rd} = 360 \left ( \frac{3M_3a_{\rm in}^3}{4(M_1 + M_2)a_{\rm out}^3}\right )  \text{ deg cycle}^{-1}
\end{equation}

where $a_{\rm out}$ is the perturber's semimajor axis and $M_3$ is the mass of the perturber. We are especially interested in the region just outside the binary's critical stability radius, given the high abundance of CBPs residing in this area. 



\begin{figure}
    \centering
    \includegraphics[width=1.0\linewidth]{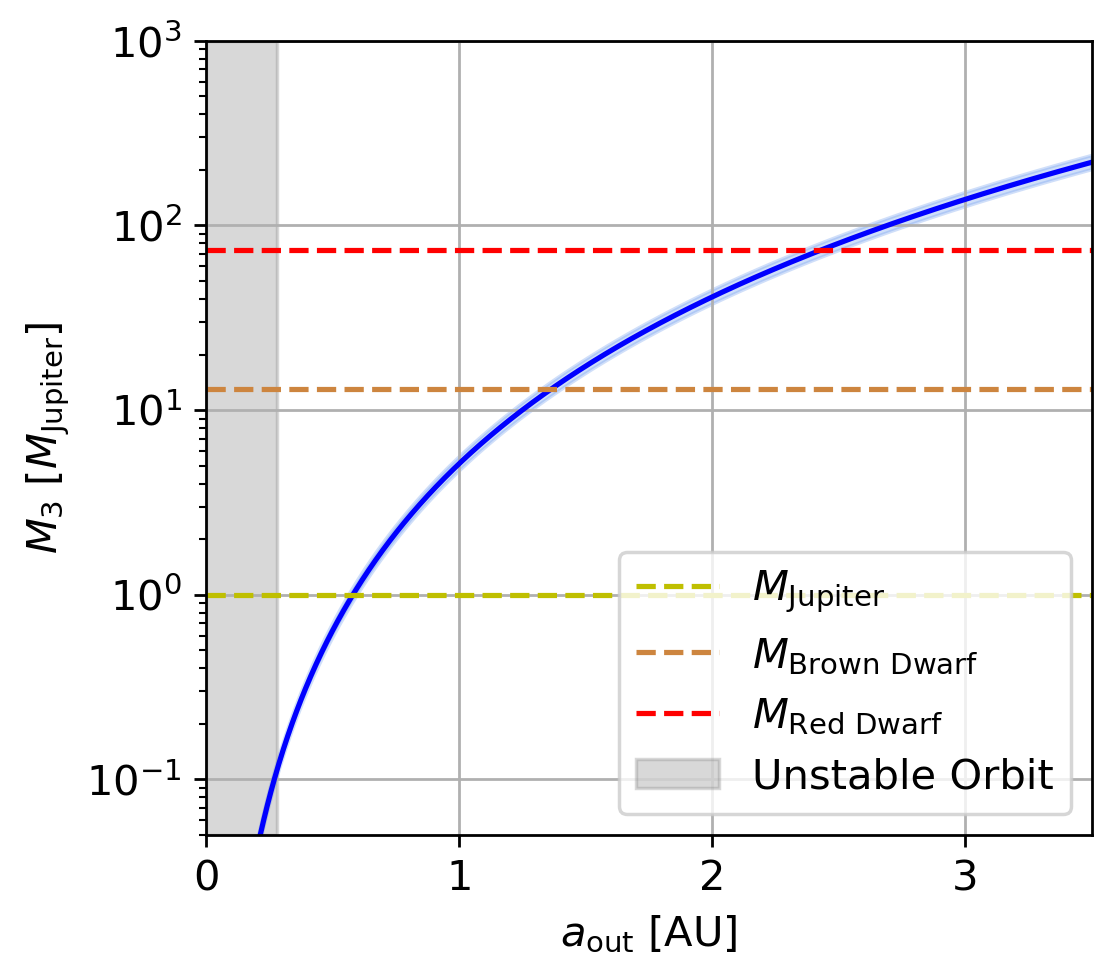}
    \caption{The parameter space for the companion to TIC 286310830, represented by the curve. This precession could be due to a $\sim0.15 M_{\rm Jupiter}$ companion at $\sim0.25$ AU. The region within the critical stability radius is shaded in on the left.}
    \label{fig:TIC286310830_3rd}
\end{figure}

The perturber could fall anywhere along the blue curve defined by Figure \ref{fig:TIC286310830_3rd} outside of the critical radius. The lower mass limit, $M_{3,\textrm{min}}$, is then determined as the point where the perturber parameter curve (blue line) intersects the critical radius (right edge of grey shaded region). The upper mass limit is poorly constrained at this point, but will be refined in future work (see Section \ref{sec:futurework} for more details).

\section{Results} \label{sec:results}

We visually inspect the O$-$C and folded light curve plots of the 1,590 remaining systems and identify 71 as potentially exhibiting evidence of apsidal precession. This initial selection is based on the presence of a systematic and coherent divergence between the primary and secondary eclipse timing residuals over time, rather than random or noise-like scatter. In particular, we search for the characteristic anti-correlated pattern in which the O$-$C curves of the primary and secondary eclipses evolve in opposite directions, as expected for apsidal motion. We do not require the signal to exceed a formal significance threshold at this stage, as our goal is to construct a candidate sample rather than a statistically complete census. This initial selection is solely to identify the targets that we analyse more carefully. After subtracting the terms $\dot\omega_{\rm GR}$ and $\dot\omega_{\rm CL}$, 36 EBs remain with precession that cannot be accounted for by GR and tidal/rotational effects, outlined in Table \ref{tab:tic_omega_dot}. 

Taken together, the 36 systems exhibiting residual apsidal precession and the 31 systems showing LTTE signatures imply that approximately 4\% of the 1,590 inspected binaries display timing variations consistent with the presence of a third body. For comparison, in a survey of the \textit{Kepler} EBs, \citep{rappaport2013} identified 39 triple-star candidates out of roughly 1000 EBs analysed, of which 15 host binaries have inner orbital periods longer than 3 days, matching the period range considered in our work. They further estimate that their detected sample represents roughly half of the true underlying population due to sensitivity and time-baseline limitations. Accounting for this incompleteness implies an effective occurrence rate of $3\%$ for triples among \textit{Kepler} EBs with $P_{\rm Binary} > 3$ days. Our inferred $\sim4\%$ rate includes contributions from both stellar and possible planetary companions; restricting our sample to triple-star systems alone yields an occurrence rate of $\sim2.5\%$, in close agreement with the estimate of \citep{rappaport2013}.


We also note 31 systems revealing evidence of the light travel time effect (LTTE); we provide the TIC ID's and semi-amplitude values of the LTTE signal in Table \ref{tab:ltte}. Otherwise known as the R\o mer Delay \citep{romer}, LTTE is the observed shift in timing of both eclipses due to a change in the light's path length to Earth as the barycentre of the binary wobbles around due to a third companion. Detectable LTTE in such cases would likely be caused by brown dwarf, stellar, or higher mass companions. Although these are interesting, the results of this paper focus on possible planet-mass companions. Thus, we focus on the 36 EBs in Table \ref{tab:tic_omega_dot}. 

TIC 84546771 and TIC 198242678 both show signals that resemble fast, shorter-term precession likely caused by a more massive companion. TIC 115396972 is likely to have non-planetary-mass companions, as the minimum perturber mass lies right on the upper limit of a brown dwarf mass. We also exclude TIC 444544588 and TIC 167756615 due to their renormalised unit weight error (RUWE) from \textit{Gaia} \citep{Gaia2021}. A high RUWE value ($> \sim 1.4$) suggests that the star's motion is not well described by a single-star model, which is often a sign of an unresolved companion inducing detectable photocentre motion. For the systems in our sample, planet-mass companions at circumbinary separations near the stability limit are expected to induce astrometric signals far below \textit{Gaia}'s detection threshold (typically tens of $\mu as$; \citealt{perryman2011, Gaia2016}). Thus, \textit{Gaia} RUWE is not sensitive to the types of CBPs we aim to detect. Likewise, the inner binaries themselves are unlikely to contribute significantly to the RUWE since the binaries we are looking at produce photocentre motions that are below \textit{Gaia}'s resolution and are effectively averaged out over the mission baseline \citep[][]{stassun2021}. The RUWE for TIC 444544588 and TIC 167756615 are 2.45 and 2.23, respectively; therefore, they are likely to have a more massive companion. Although TIC 441496809, TIC 165615442, TIC 121092916, and TIC 342356517 exhibit apsidal precession consistent with a potential planetary-mass tertiary companion, their measured precession rates are all below a $3\sigma$ significance threshold. As such, we do not consider them candidates at this stage. Additional eclipse timing is required to confirm whether these systems host third bodies; for that reason, we exclude them from our candidate count. We therefore present 27 candidate circumbinary planets, with more than half permitting the existence of sub-Jupiter mass objects. 

\begin{table*}
\vfill
\centering
\begin{tabular}{ccccccc}
\hline
TIC ID & $P$ & $\dot{\omega}_{\rm obs}$  & $\dot{\omega}_{\rm GR}$ & $\dot{\omega}_{\rm CL}$ & $\dot{\omega}_{\rm 3rd}$ & $M_{3,\textrm{min}}$\\
 & [days] & [$10^{-3}$ deg/cycle]  & [$ 10^{-3}$ deg/cycle] & [$10^{-3}$ deg/cycle] & [$10^{-3}$ deg/cycle] & [$M_{\textrm{Jup}}$]\\
\hline
123951716 & $23.1065 \pm 0.0003$ & $3.84 \pm 0.01$ & $0.136 \pm 0.006$ & $0.003 \pm 0.002$ & $3.7 \pm 0.1$ & $0.14 \pm 0.01
$\\
81741369 & $10.0920 \pm 0.0001$ & $0.85 \pm 0.01$ & $0.253 \pm 0.009$ & $0.02 \pm 0.01$ & $0.58 \pm 0.02$ & $0.101 \pm 0.003$\\
293225466 & $129.3214 \pm 0.0008$ & $13.3 \pm 0.7$ & $0.036 \pm 0.002$ & $0.000 \pm 0.000$ & $13.3 \pm 0.7$ & $1.43 \pm 0.08$\\
457579488 & $14.6925 \pm 0.0005$ & $0.48 \pm 0.02$ & $0.182 \pm 0.008$ & $0.01 \pm 0.01$ & $0.28 \pm 0.02$ & $0.038 \pm 0.003$\\
158330804$^1$ & $11.355 \pm 0.002$ & $0.61 \pm 0.03$ & $0.19 \pm 0.01$ & $0.032 \pm 0.001$ & $0.38 \pm 0.03$ & $0.055 \pm 0.005$\\
286310830 & $8.3687 \pm 0.0003$ & $1.12 \pm 0.07$ & $0.22 \pm 0.01$ & $0.02 \pm 0.02$ & $0.88 \pm 0.07$ & $0.109 \pm 0.009$\\
274182408 & $6.554 \pm 0.001$ & $1.9 \pm 0.1$ & $0.26 \pm 0.01$ & $0.06 \pm 0.05$ & $1.6 \pm 0.1$ & $0.19 \pm 0.01$\\
292357653 & $23.1476 \pm 0.0002$ & $0.507 \pm 0.009$ & $0.15 \pm 0.01$ & $0.02 \pm 0.03$ & $0.34 \pm 0.03$ & $0.040 \pm 0.004$\\
124282654 & $11.3293 \pm 0.0002$ & $2.9 \pm 0.3$ & $0.172 \pm 0.008$ & $0.008 \pm 0.008$ & $2.7 \pm 0.3$ & $0.29 \pm 0.03$\\
51629874 & $6.465 \pm 0.001$ & $14.7 \pm 1.4$ & $0.23 \pm 0.04$ & $0.04 \pm 0.04$ & $14.4 \pm 1.4$ & $1.5 \pm 0.1$\\
396170777 & $6.5656 \pm 0.0002$ & $1.3 \pm 0.1$ & $0.25 \pm 0.01$ & $0.05 \pm 0.04$ & $1.0 \pm 0.1$ & $0.12 \pm 0.02$\\
56023695 & $8.177 \pm 0.002$ & $27.3 \pm 2.7$ & $0.39 \pm 0.01$ & $1.2 \pm 0.7$ & $25.7 \pm 2.8$ & $5.0 \pm 0.5$\\
253270207 & $6.106 \pm 0.006$ & $5.1 \pm 0.5$ & $0.34 \pm 0.01$ & $0.10 \pm 0.09$ & $4.7 \pm 0.5$ & $0.70 \pm 0.07$\\
82893635 & $28.5504 \pm 0.0007$ & $0.76 \pm 0.07$ & $0.13 \pm 0.01$ & $0.01 \pm 0.01$ & $0.61 \pm 0.07$ & $0.084 \pm 0.009$\\
146204045$^1$ & $14.192 \pm 0.001$ & $7.2 \pm 0.8$ & $0.205 \pm 0.006$ & $0.005 \pm 0.003$ & $7.0 \pm 0.8$ & $1.3 \pm 0.2$\\
283651681 & $7.622 \pm 0.009$ & $1.1 \pm 0.1$ & $0.24 \pm 0.01$ & $0.03 \pm 0.03$ & $0.8 \pm 0.1$ & $0.11 \pm 0.02$\\
61656788   & $4.3028 \pm 0.0007$ & $1.79 \pm 0.03$ & $0.42 \pm 0.01$ & $0.2 \pm 0.1$ & $1.1 \pm 0.1$ & $0.19 \pm 0.02$\\
444544588 & $16.349 \pm 0.002$ & $24.3 \pm 3.1$ & $0.25 \pm 0.01$ & $0.11 \pm 0.06$ & $23.9 \pm 3.1$ & $4.8 \pm 0.6$\\
139699256$^1$ & $5.952 \pm 0.001$ & $1.3 \pm 0.1$ & $0.29 \pm 0.01$ & $0.07 \pm 0.04$ & $1.0 \pm 0.1$ & $0.14 \pm 0.02$\\
252497283 & $5.0403 \pm 0.0003$ & $2.7 \pm 0.3$ & $0.30 \pm 0.01$ & $0.1 \pm 0.1$ & $2.3 \pm 0.3$  & $0.28 \pm 0.04$\\
196989952 & $7.281 \pm 0.002$ & $2.57 \pm 0.03$ & $0.42 \pm 0.01$ & $0.4 \pm 0.2$ & $1.7 \pm 0.2$ & $0.27 \pm 0.04$\\
167699456 & $18.743 \pm 0.007$ & $4.0 \pm 0.6$ & $0.11 \pm 0.01$ & $0.003 \pm 0.008$ & $3.9 \pm 0.6$ & $0.39 \pm 0.06$\\
343127696 & $4.672 \pm 0.002$ & $22.2 \pm 3.5$ & $0.31 \pm 0.01$ & $0.2 \pm 0.1$ & $21.7 \pm 3.5$ & $2.4 \pm 0.4$\\
115396972 & $19.809 \pm 0.001$ & $64 \pm 17$ & $0.18 \pm 0.01$ & $0.03 \pm 0.02$ & $64 \pm 17$ & $ 10.0 \pm 0.2$\\
399127035 & $5.448 \pm 0.001$ & $8.8 \pm 2.4$ & $0.26 \pm 0.01$ & $0.1 \pm 0.1$ & $8.5 \pm 2.4$ & $0.9 \pm 0.3$\\
355503224 & $6.9247 \pm 0.0002$ & $0.8 \pm 0.1$ & $0.26 \pm 0.01$ & $0.04 \pm 0.02$ & $0.5 \pm 0.1$ & $0.07 \pm 0.02$\\
95622298 & $4.5123 \pm 0.0003$ & $5.9 \pm 1.6$ & $0.32 \pm 0.01$ & $0.1 \pm 0.1$ & $5.4 \pm 1.6$ & $0.6 \pm 0.2$\\
64366964$^1$ & $4.459 \pm 0.003$ & $11.5 \pm 0.7$ & $0.49 \pm 0.01$ & $5.9 \pm 1.5$ & $5.1 \pm 1.6$ & $0.8 \pm 0.3$\\
291751499 & $5.1957 \pm 0.0003$ & $16.2 \pm 5.2$ & $0.30 \pm 0.01$ & $0.11 \pm 0.08$ & $15.8 \pm 5.2$ & $1.8 \pm 0.8$\\ 
198242678 & $4.6298 \pm 0.0009$ & $65 \pm 22$ & $0.27 \pm 0.01$ & $0.3 \pm 0.3$ & $65 \pm 22$ & $6.1 \pm 2.0$\\
84546771 & $4.2837 \pm 0.0009$ & $42 \pm 14$ & $0.37 \pm 0.01$ & $0.2 \pm 0.1$ & $41 \pm 14$ & $5.7 \pm 1.7$\\
167756615 & $19.181 \pm 0.007$ & $15.8 \pm 5.8$ & $0.12 \pm 0.01$ & $0.001 \pm 0.001$ & $15.7 \pm 5.8$ & $1.8 \pm 0.6$\\
441496809 & $5.7542 \pm 0.0008$ & $4.8 \pm 1.9$ & $0.26 \pm 0.01$ & $0.08 \pm 0.09$ & $4.5 \pm 1.9$ & $0.5 \pm 0.2$\\
165615442 & $5.475 \pm 0.001$ & $13.0 \pm 6.0$ & $0.27 \pm 0.01$ & $0.08 \pm 0.09$ & $12.7 \pm 6.0$ & $1.3 \pm 0.6$\\
121092916 & $8.2198 \pm 0.0002$ & $7.0 \pm 4.3$ & $0.21 \pm 0.01$ & $0.02 \pm 0.02$ & $6.8 \pm 4.3$ & $0.7 \pm 0.4$\\
342356517 & $11.9666 \pm 0.0009$ & $1.9 \pm 1.4$ & $0.158 \pm 0.008$ & $0.008 \pm 0.009$ & $1.8 \pm 1.4$ & $0.2 \pm 0.1$\\
\hline
\multicolumn{7}{l}{$^1$ Systems which also exhibit the LTTE.}\\
\end{tabular}
\vfill
\caption{Period, apsidal precession contributions, and minimum perturber mass for all precessing systems. Rows are sorted by descending $\dot{\omega}_{\rm 3rd}/\sigma_{\dot{\omega}_{\rm 3rd}}$. The period is the average of the primary and secondary period.}
\label{tab:tic_omega_dot}
\end{table*}

The O$-$C plots for the systems from Table \ref{tab:tic_omega_dot} are shown in Figures \ref{fig:oc_plots1} - \ref{fig:TIC198242678_OC}, where Figures \ref{fig:TIC84546771_OC_cut} and \ref{fig:TIC198242678_OC} show the systems with short-term variations in the O$-$C plot. The mass vs. semimajor axis plots are shown in Figures \ref{fig:MA_plots1} - \ref{fig:MA_plots3}. 

We test the methods on TIC 172900988, the EB discussed above with a known CBP and observed apsidal precession. Our analysis resulted in a $\dot{\omega}_{\rm 3rd}$ of 2.41 $\pm$ 0.11 $\times 10^{-3}$ degrees per cycle compared to the estimate of 2.71 $\times 10^{-3}$ degrees per cycle from \citet{TIC1729}. Considering the full apsidal period is $\sim 50$ years, we expect a $\sim10\%$ discrepancy between the two calculations due to a change in phase of $\cos \omega$ between these observations and the 2006-2021 observations used in that paper. This test validates the methods used to detect and quantify apsidal precession.

\begin{table}
\centering
\begin{tabular}{ccc}
\hline
TIC ID & Semi-Amplitude [min] & Period [days] \\
\hline
280625073 & 6.9 $\pm$ 1.9 & 3514 $\pm$ 691 \\
243588685 & 5.7 $\pm$ 2.3 & 1455 $\pm$ 10 \\
360088090 & 1.5 $\pm$ 0.5 & 1879 $\pm$ 758 \\
445259184 & 2.9 $\pm$ 0.8 & 2732 $\pm$ 992 \\
9655156   & 7.3 $\pm$ 1.0 & 1645 $\pm$ 31 \\
225138160 & 3.6 $\pm$ 1.0 & 1394 $\pm$ 37 \\
447423507 & 0.7 $\pm$ 0.1 & 1322 $\pm$ 193 \\
279638469 & 3.9 $\pm$ 1.3 & 3022 $\pm$ 720 \\
407027744 & 1.9 $\pm$ 0.1 & 1650 $\pm$ 13 \\
361887558 & 2.3 $\pm$ 0.6 & 1256 $\pm$ 220 \\
301260701 & 3.0 $\pm$ 0.7 & 1407 $\pm$ 133 \\
81923827  & 1.0 $\pm$ 0.4 & 1828 $\pm$ 226 \\
276471941 & 5.4 $\pm$ 2.3 & 1277 $\pm$ 84 \\
265338722 & 2.4 $\pm$ 0.1 & 1915 $\pm$ 38 \\
160676497 & 3.8 $\pm$ 1.2 & 1156 $\pm$ 219 \\
236212010 & 1.4 $\pm$ 0.2 & 1022 $\pm$ 161 \\
293271277 & 3.2 $\pm$ 0.7 & 4000 $\pm$ 632 \\
418011373 & 6.0 $\pm$ 3.1 & 1531 $\pm$ 44 \\
245217638 & 4.8 $\pm$ 0.2 & 2434 $\pm$ 828 \\
64355437  & 3.0 $\pm$ 1.0 & 1116 $\pm$ 9 \\
165295354 & 1.7 $\pm$ 0.1 & 2100 $\pm$ 106 \\
265288713 & 1.7 $\pm$ 0.9 & 1850 $\pm$ 23 \\
436131579 & 3.7 $\pm$ 1.1 & 1387 $\pm$ 139 \\
359960689 & 5.8 $\pm$ 0.9 & 2456 $\pm$ 419 \\
279647970 & 1.3 $\pm$ 0.2 & 1810 $\pm$ 21 \\
407165593 & 3.6 $\pm$ 0.2 & 1722 $\pm$ 447 \\
45637974  & 5.5 $\pm$ 0.6 & 1544 $\pm$ 11 \\
144215102 & 2.3 $\pm$ 0.2 & 1921 $\pm$ 87 \\
311706965 & 4.3 $\pm$ 0.8 & 1737 $\pm$ 41 \\
220113185 & 2.6 $\pm$ 0.6 & 2435 $\pm$ 619 \\
157089542 & 1.0 $\pm$ 0.1 & 2174 $\pm$ 38 \\
158330804 & 0.5 $\pm$ 0.2 & 6125 $\pm$ 954 \\
139699256 & 2.0 $\pm$ 0.4 & 10056 $\pm$ 1020 \\
64366964  & 3.2 $\pm$ 1.0 & 3506 $\pm$ 690 \\
146204045 & 16.8 $\pm$ 0.1 & 4803 $\pm$ 64 \\
\hline
\end{tabular}
\caption{The eclipsing binaries that exhibit signs of the LTTE. The semi-amplitudes and periods and their corresponding uncertainties are obtained by fitting a circular orbit (sinusoid) to the signal. Follow-up modelling and data are required for these systems. The last five systems also exhibit precession signals; the observed precession of TIC 157089542 can be fully accounted for by general relativity and tides, whereas the remaining four likely host an additional perturber.}
\label{tab:ltte}
\end{table}

%


\section{Discussion} \label{sec:discussion}

In this work, we demonstrate that we can detect apsidal precession of EBs at a level of precision that enables the detection of sub-Jovian mass planets.  We identify 27 systems that could host a planetary mass perturber, with over half allowing for sub-Jupiter mass. If the general population follows trends of the known CBPs, we can infer that the tertiary companions are likely to exist just outside of the critical radius of the binary. In this case, all 27 systems could be confirmed as CBPs. Refining the mass to rule out higher-mass companions is required for confirmation and relies on attaining RV data (see Section \ref{sec:futurework} for more information). 

\subsection{Comparison to the Known CBP Population}

The 27 candidates identified in this work differ in several ways from the currently known CBPs discovered via transits. Transit surveys preferentially detect systems with cool, low-luminosity primaries, long eclipse durations, and deep transits; these biases make CBP discoveries strongly skewed towards Sun-like or cooler binaries. In contrast, our precession-based method is largely insensitive to stellar radius, luminosity, or transit depth. As a result, our sample includes a broader range of stellar types, including a fraction of hotter and larger systems for which transit detection would be extremely challenging. There are no known CBPs discovered by the transit method orbiting stars hotter than 6400 K, while 30\% of our candidates are: the hottest stars in our sample have effective temperatures over 9000 K. Despite this, rapid rotation in these hotter stars is unlikely to produce spurious apsidal precession signals, as only 3 of the 36 precessing systems have both $_{T_{\rm eff}} > 6300$ K and a tidal precession rate that is a non-negligible fraction ($> 10\%$) of the observed apsidal precession. Nevertheless, the role of tidal locking in these hotter stars is poorly constrained, and future measurements of stellar rotation ($v \sin i$) will be important for directly assessing the influence of rotation on tidal effects in these systems. Additionally, the EBs in our sample have a wider range of orbital periods than the EBs with known CBPs: our shortest system has an orbital period of $4.2837 \pm 0.0009$ days, while our longest has a period of $129.3214 \pm 0.0008$ days, both outside of the range of periods of confirmed transiting planet-hosting CBPs. A full demographic comparison (distributions of perturber mass, inclination, etc.) is beyond the scope of this work but will be presented in future analyses.

\subsection{LTTE and Precession}

TIC 158330804, TIC 139699256, TIC 64366964, and TIC 146204045 show signs of LTTE in addition to the diverging precession signal in their O$-$C diagrams (see Table \ref{tab:ltte} for semi-amplitude estimates). This double effect may be particularly informative, because the combination of apsidal precession and LTTE provides complementary constraints that can help break degeneracies in the inferred mass and orbital configuration of the tertiary companion. In principle, LTTE and apsidal precession could be produced by the same object or by two independent companions. Wider-baseline data or dynamical modelling is ultimately required to confirm or rule out multi-perturber configurations.

\subsection{Short Term Variation}

TIC 84546771 and TIC 198242678 both show short-term variations that require more follow-up to determine the cause of such precession. Possible causes of these signals include a very massive or nearby object. In both cases, the rapid evolution of the O$–$C curve suggests a dynamical influence that cannot be fully characterised with the current data. Continued timing monitoring and spectroscopic follow-up will be essential to determine whether these variations arise from a compact tertiary companion, dynamical interactions within a hierarchical system, or an alternative solution.

Short-term apsidal variation of this kind can be produced by (1) a massive tertiary on a moderately eccentric orbit, (2) resonant hierarchical architectures, or (3) combinations of multiple low-mass perturbers whose interactions are nonlinear over the \textit{TESS} baseline. These systems are therefore strong candidates for full N-body dynamical modelling, but such work requires more precise constraints on the binary parameters. Future RV measurements and extended eclipse timing baselines are essential to determine whether these are unusual CBP configurations or more complex hierarchical triples.

\subsection{Low-Mass Cases}

The methods developed here allow for the detection of tertiary companions down to sub-Neptune masses. For several systems, specifically TIC 457579488, TIC 292357653, TIC 355503224, and TIC 158330804, the intersection of the perturbers allowed parameter space with the system’s critical radius permits solutions involving companions less massive than Neptune. A circumbinary planet of this mass would be the least massive detected to date, placing these systems at the frontier of current sensitivity limits. Continued monitoring and complementary follow-up observations are crucial for confirming such low-mass candidates and refining their orbital properties.

A natural question is whether these low-mass candidates are detectable through radial velocities alone. For a sub-Neptune mass CBP orbiting just outside the critical radius (typically 0.3-0.6 AU), the reflex velocity amplitude of the binary barycentre is not more than a few m\,s$^{-1}$, even if the system is edge on. This magnitude places these planets near the limit of current state-of-the-art precision for surveys of bright systems such as BEBOP \citep{bebop, standing2022, triaud2022}. Thus, RVs may be able to place meaningful upper limits, but unambiguous planet detection may require a combination of long-baseline ETV monitoring, dynamical modelling, and astrometric constraints from future \textit{Gaia} releases. In particular, while current \textit{Gaia} data primarily provide summary diagnostics of non-single-star behaviour (e.g. RUWE), future data products will enable direct orbital modelling of tertiary companions, substantially improving sensitivity to low-mass and wide companions.

\subsection{Mass–Separation Limits and Companion Types}

Apsidal precession places limits on the mass–semimajor axis relationship of permitted companions, with the required mass to induce the observed precession growing as $a^3$. Stellar tertiaries beyond a few AU would, if sufficiently massive, outshine the binary and be easily identified spectroscopically. \textit{Gaia} astrometry already constrains the presence of stellar or brown-dwarf companions for many systems, and \textit{Gaia} DR4 will significantly tighten these limits. 

While our analysis assumes main-sequence tertiary companions, compact objects (white dwarfs, neutron stars, black holes) could in principle produce similar dynamical signatures. Most such companions at the relevant separations would produce detectable RV accelerations or \textit{Gaia} astrometric motion. With the availability of epoch astrometry and RVs in \textit{Gaia} DR4, it will become possible to directly test these scenarios on a system-by-system basis. Future observations will help determine whether any observed precession is caused by such compact objects.

\subsection{Future Work} \label{sec:futurework}

The 27 CBP candidates require follow-up radial velocity observations to refine their mass estimates and to rule out brown dwarf or stellar-mass tertiaries. All candidates are bright enough (T $<$ 13.5) for precise RV follow-up with modern 4–10 m class spectrographs. These systems therefore represent promising targets for future observing campaigns rather than relying on specific facility commitments.

Considering the sample we analysed was only a small fraction of the 2 million EBs in \textit{Gaia}'s catalogue \citep{Gaia}, we anticipate many more detections as we expand this search across the larger sample. Additionally, combining the 7 year \textit{TESS} data with past and future missions will increase the baseline of the search, and thus the sensitivity to small companions. Longer baselines will not only improve constraints on apsidal precession but also allow for the identification of multi-perturber systems and time-variable dynamical regimes that are currently indistinguishable within the existing data. Also, an in-depth transit search may find companions in these systems, making follow-up more straightforward. The combination of RVs, eclipse timing, \textit{Gaia} epoch astrometry with DR4, and systematic transit searches has the potential to transform the circumbinary companion population from a handful of transiting planets to a statistically meaningful, dynamically rich sample.

\section{Conclusions} \label{sec:conclusion}

Upon initial inspection of 1,590 eclipsing binaries from \textit{Gaia}'s catalogue of over 2 million systems, we present 71 EBs that show signs of precession, 36 of which cannot be accounted for by effects due to general relativity and the rotational/tidal effects of the stars themselves. The calculation of precession was based on the change in the argument of periastron over time of the binary star, which can be determined by the exact timing of both primary and secondary eclipses. 27 of the 36 EBs may host a planet-mass companion in an outer orbit. We therefore present the discovery of 27 circumbinary planet candidates that induce apsidal precession on their host binary. Once further analysis is conducted to determine the perturber mass, we can confirm them as CBPs. 

This method allows for the possibility of multiplying the number of known circumbinary planets because it is not limited in the orientation of the planet. It has the capability of detecting CBPs with higher mutual inclinations, therefore extending the sample to diverse architectures and orbital geometries. The findings of this work will allow us to robustly test formation theories, constrain migration histories, and understand long-term evolution of circumbinary systems.

\section*{Acknowledgements}

We wish to express our gratitude to the \textit{TESS} Mission team, the Quick-Look Pipeline developers, and the \textit{Gaia} project scientists and operations teams, whose dedicated efforts in producing and maintaining high-quality photometric and astrometric data have made this work possible. We are also grateful to our collaborators for their willingness to support and contribute to the follow-up observations associated with this project.

This paper includes data collected by the \textit{TESS} mission and processed by the Science Processing Operations Center. Funding for the \textit{TESS} mission is provided by the NASA Science Mission Directorate. This research has made use of data products from the \textit{Gaia} mission, funded by the European Space Agency (ESA), and of resources hosted at the Mikulski Archive for Space Telescopes (MAST), operated by the Space Telescope Science Institute under NASA contract NAS5-26555.

M.T. acknowledges support from the University of New South Wales through the University International Postgraduate Award and the Burbage Astronomy Student Fund.

Finally, we thank the anonymous referee for their careful reading of the paper and for providing insightful comments that improved the clarity and presentation of this work.

\section*{Data Availability}

The data underlying this article are publicly available. \textit{TESS} light curves and associated data products can be accessed through MAST. \textit{Gaia} astrometric and photometric data are available through the ESA \textit{Gaia} Archive. All additional data or analysis products used in this study will be shared upon reasonable request to the corresponding author.



\bibliographystyle{mnras}


\appendix

\section{Supplemental Figures}

\begin{figure*}
    \centering
    \includegraphics[width=1.0\linewidth]{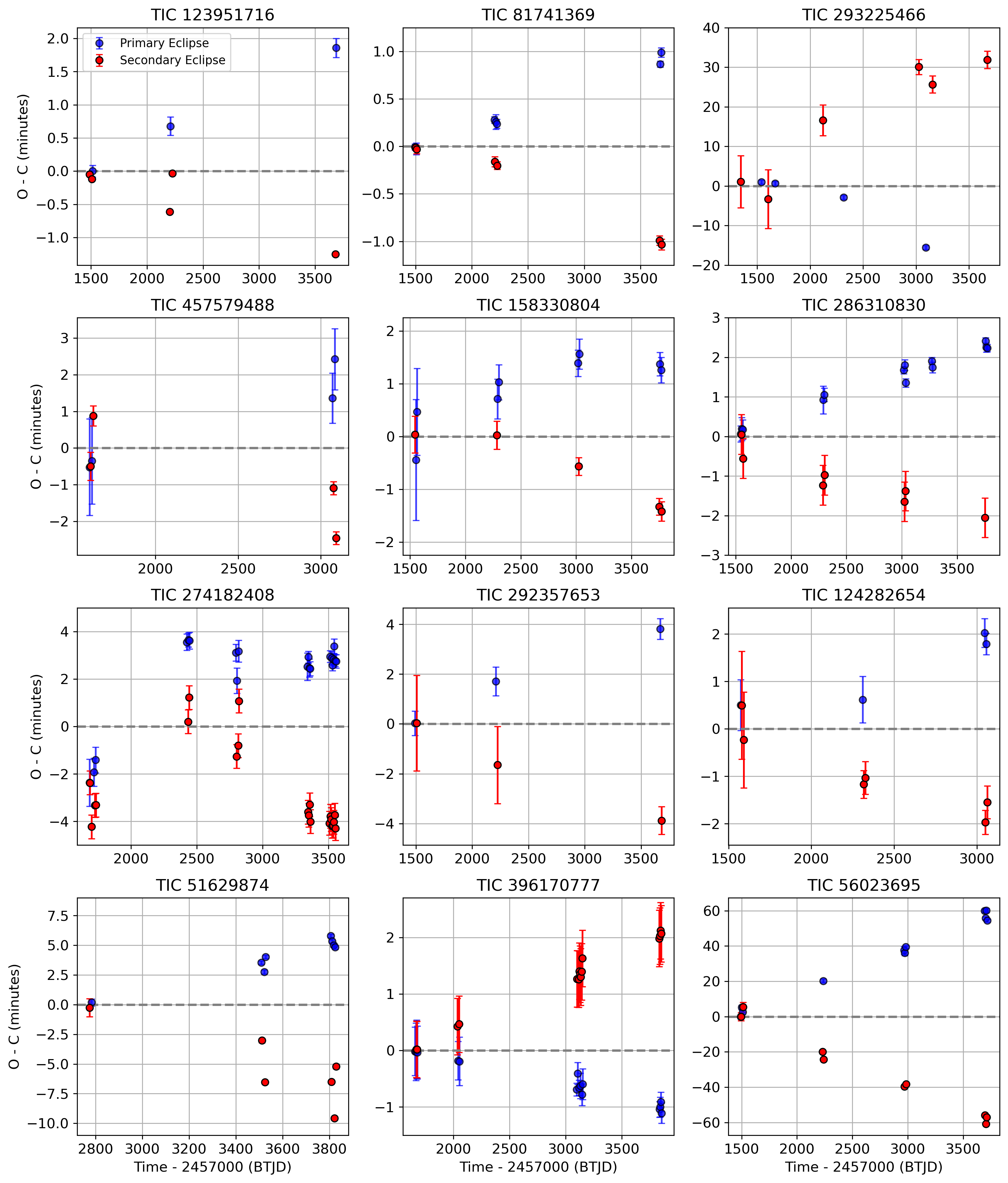}
    \caption{Same as Figure \ref{fig:TIC286310830_OCwErr}, but for the first dozen systems in Table \ref{tab:tic_omega_dot}.}
    \label{fig:oc_plots1}
\end{figure*}

\begin{figure*}
    \centering
    \includegraphics[width=1.0\linewidth]{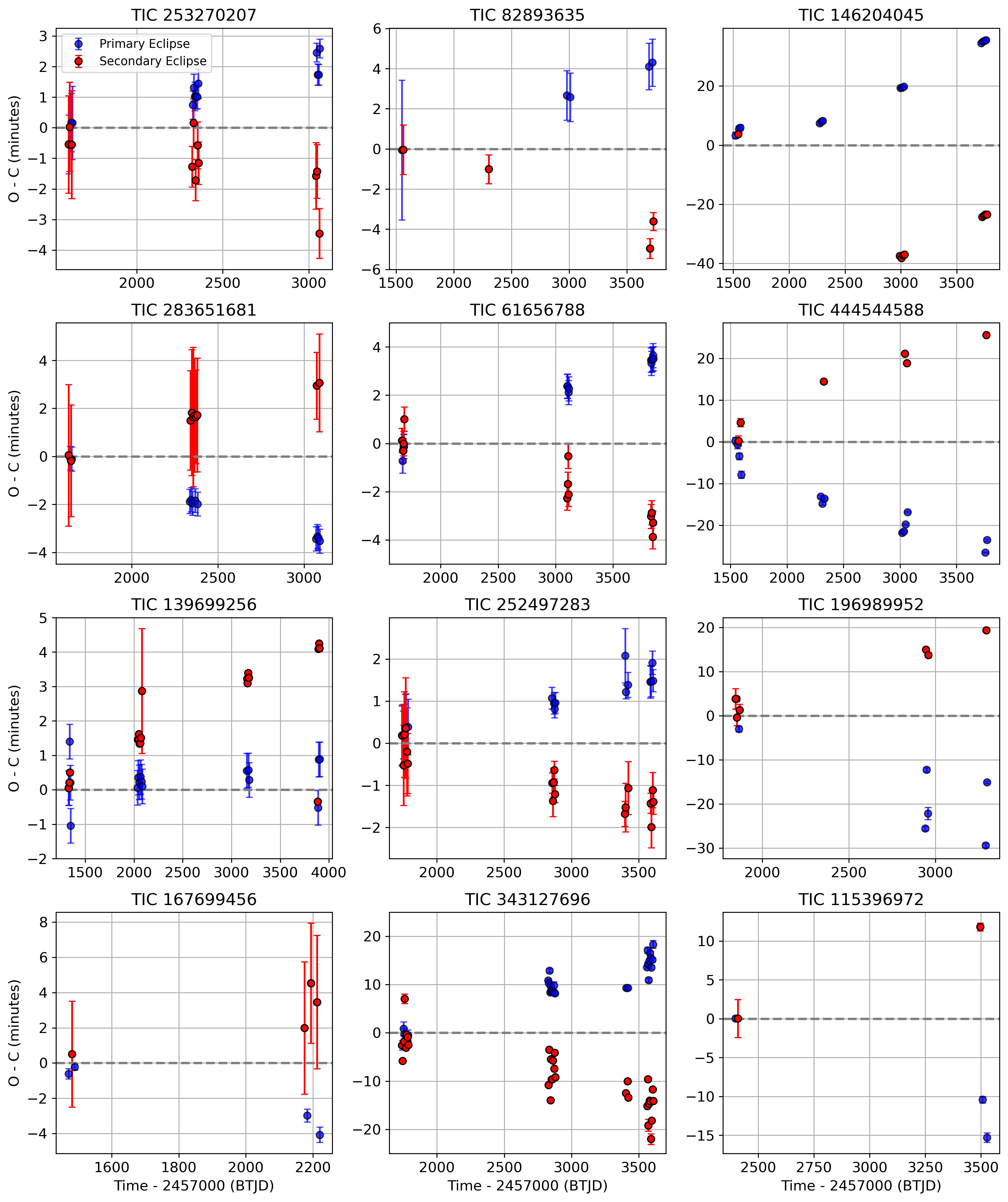}
    \caption{Same as Figure \ref{fig:TIC286310830_OCwErr}, but for the second dozen systems in Table \ref{tab:tic_omega_dot}.}
    \label{fig:oc_plots2}
\end{figure*}

\begin{figure*}
    \centering
    \includegraphics[width=1.0\linewidth]{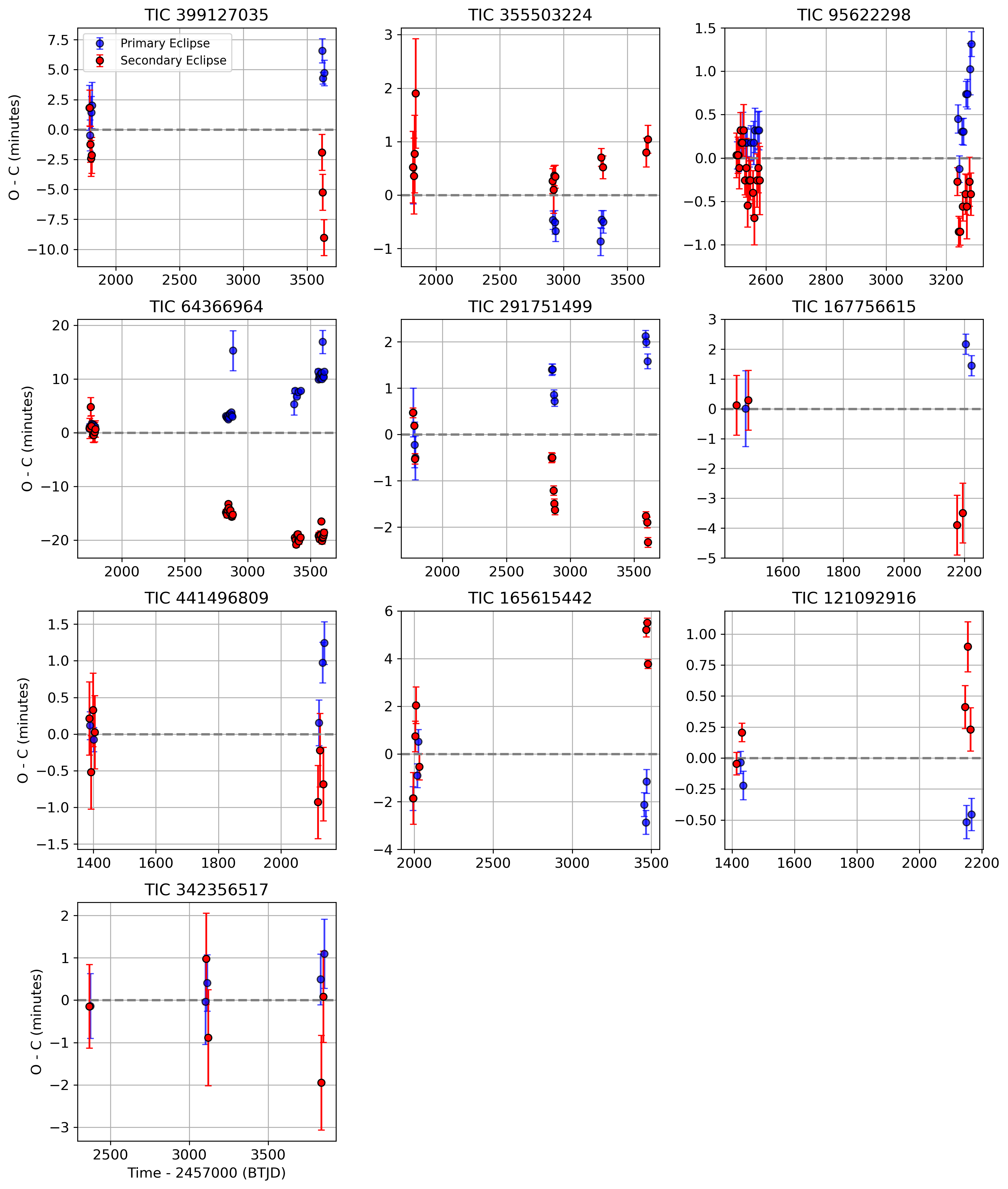}
    \caption{Same as Figure \ref{fig:TIC286310830_OCwErr}, but for the remaining systems in Table \ref{tab:tic_omega_dot}, excluding two that show short-term variation, displayed in Figures \ref{fig:TIC84546771_OC_cut} and \ref{fig:TIC198242678_OC}.}
    \label{fig:oc_plots3}
\end{figure*}

\begin{figure*}
    \centering
    \includegraphics[width=0.8\linewidth]{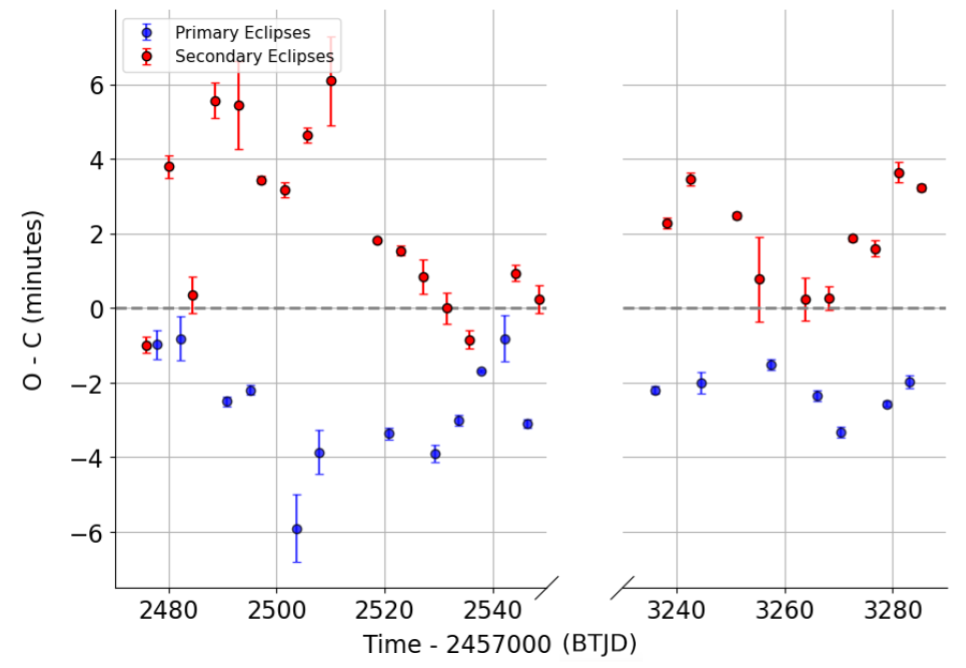}
    \caption{O-C plot of TIC 84546771, a $\sim4.28$-day EB showing short-term variation.}
    \label{fig:TIC84546771_OC_cut}
\end{figure*}

\begin{figure*}
    \centering
    \includegraphics[width=0.8\linewidth]{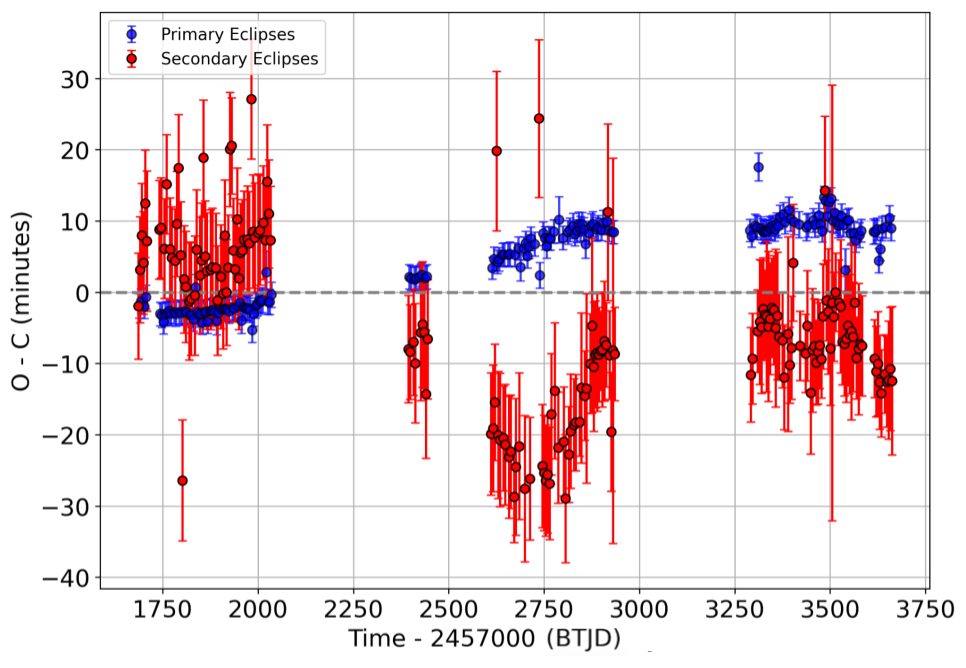}
    \caption{O-C plot of TIC 198242678, a $\sim4.63$-day EB showing short-term variation.}
    \label{fig:TIC198242678_OC}
\end{figure*}

\begin{figure*}
    \centering
    \includegraphics[width=1.0\linewidth]{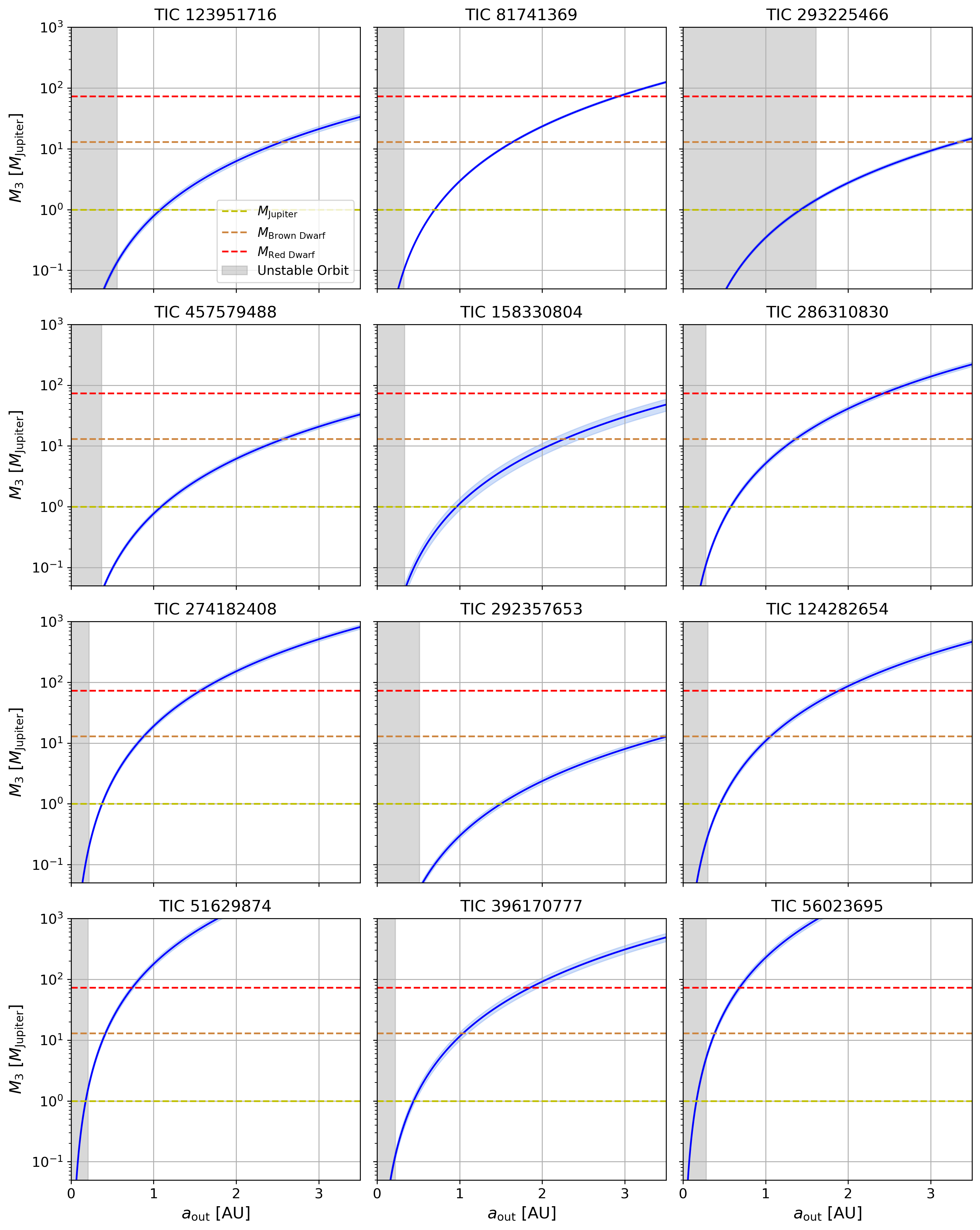}
    \caption{Same as Figure \ref{fig:TIC286310830_3rd}, but for the first dozen systems in Table \ref{tab:tic_omega_dot}.}
    \label{fig:MA_plots1}
\end{figure*}

\begin{figure*}
    \centering
    \includegraphics[width=1.0\linewidth]{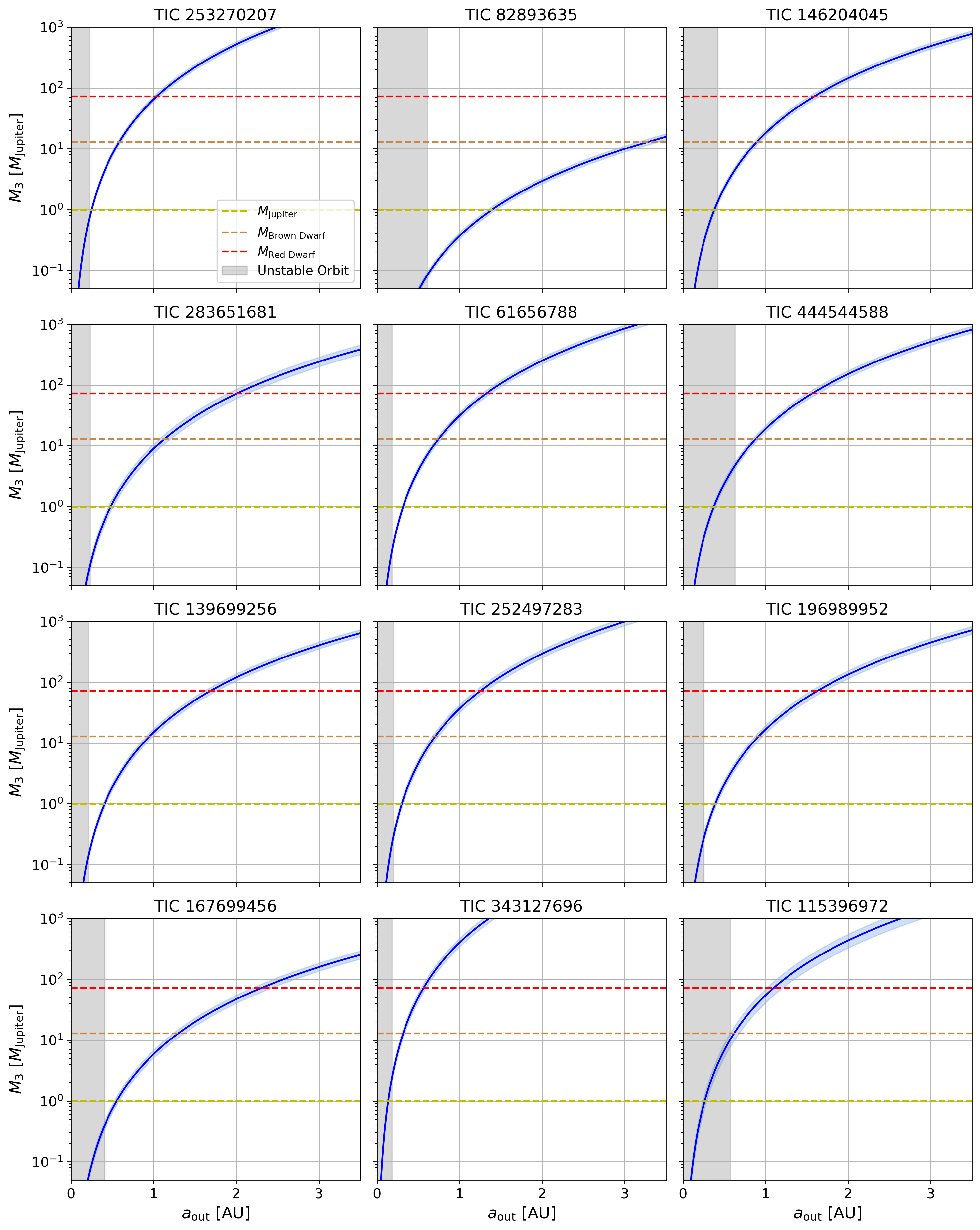}
    \caption{Same as Figure \ref{fig:TIC286310830_3rd}, but for the second dozen systems in Table \ref{tab:tic_omega_dot}.}
    \label{fig:MA_plots2}
\end{figure*}

\begin{figure*}
    \centering
    \includegraphics[width=1.0\linewidth]{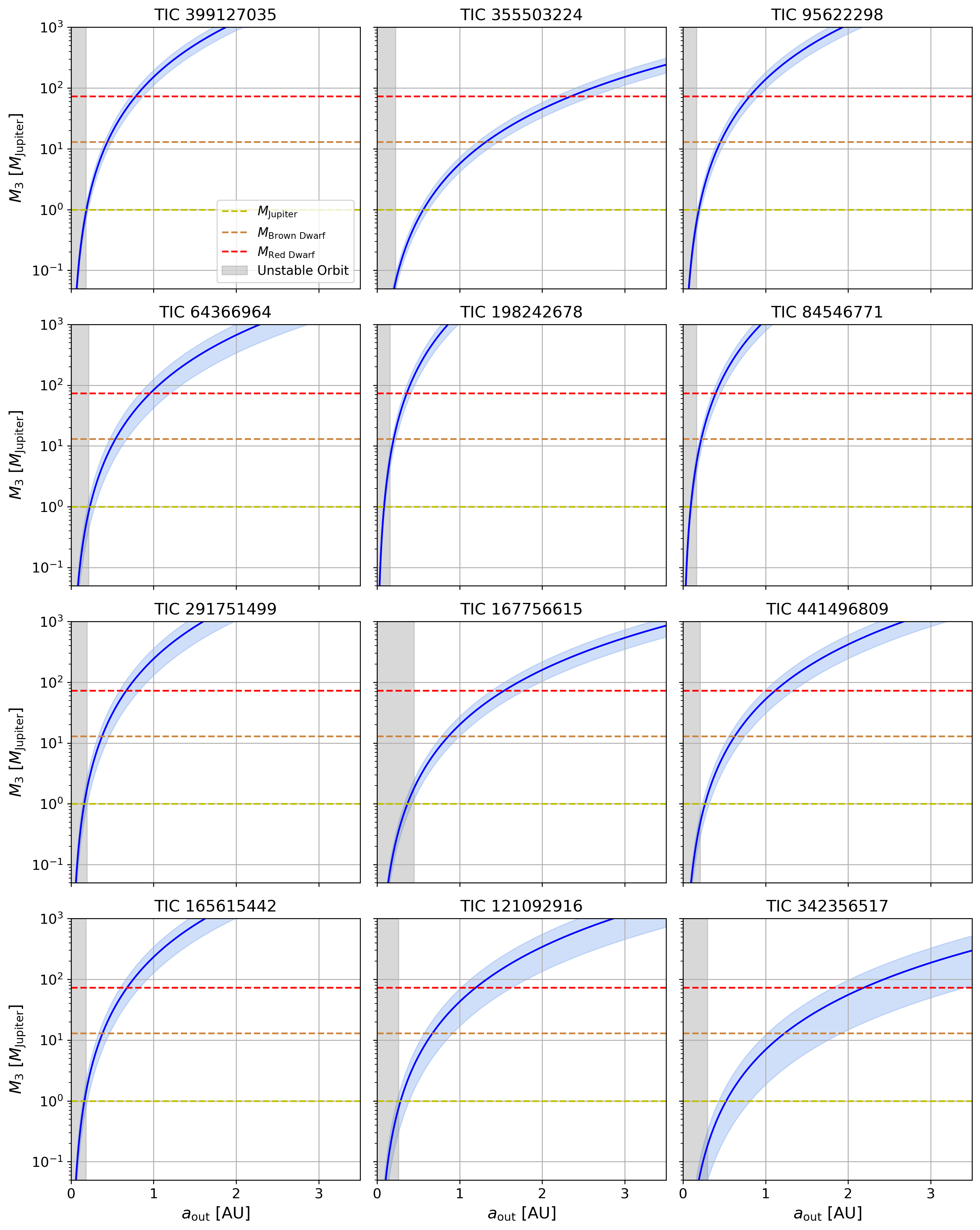}
    \caption{Same as Figure \ref{fig:TIC286310830_3rd}, but for the last dozen systems in Table \ref{tab:tic_omega_dot}.}
    \label{fig:MA_plots3}
\end{figure*}



\bsp	
\label{lastpage}
\end{document}